\documentstyle[aps,eqsecnum,preprint,floats,epsf,epsfig]{revtex}
\begin{document}
\def\be{\begin{eqnarray}}
\def\en{\end{eqnarray}}
\def\non{\nonumber}
\def\la{\langle}
\def\ra{\rangle}
\def\nc{N_c^{\rm eff}}
\def\vp{\varepsilon}
\def\drho{\bar\rho}
\def\deta{\bar\eta}
\def\vma{{_{V-A}}}
\def\vpa{{_{V+A}}}
\def\J{{J/\psi}}
\def\ov{\overline}
\def\Lqcd{{\Lambda_{\rm QCD}}}
\def\pr{{\sl Phys. Rev.}~}
\def\prl{{\sl Phys. Rev. Lett.}~}
\def\pl{{\sl Phys. Lett.}~}
\def\np{{\sl Nucl. Phys.}~}
\def\zp{{\sl Z. Phys.}~}
\def\lsim{ {\ \lower-1.2pt\vbox{\hbox{\rlap{$<$}\lower5pt\vbox{\hbox{$\sim$}
}}}\ } }
\def\gsim{ {\ \lower-1.2pt\vbox{\hbox{\rlap{$>$}\lower5pt\vbox{\hbox{$\sim$}
}}}\ } }

\font\el=cmbx10 scaled \magstep2{\obeylines \hfill May, 2001}

\vskip 1.5 cm

\centerline{\large\bf Analysis of $B\to \phi
 K$ Decays in QCD Factorization}
\bigskip
\centerline{\bf Hai-Yang Cheng$^{1,2}$ and Kwei-Chou Yang$^{3}$}
\medskip
\centerline{$^1$ Institute of Physics, Academia Sinica}
\centerline{Taipei, Taiwan 115, Republic of China}
\medskip
\centerline{$^2$ Physics Department, Brookhaven National
Laboratory} \centerline{Upton, New York 11973}
\medskip
\centerline{$^3$ Department of Physics, Chung Yuan Christian
University} \centerline{Chung-Li, Taiwan 320, Republic of China}
\bigskip
\bigskip
\centerline{\bf Abstract}
\bigskip
{\small We analyze the decay $B\to \phi K$ within the framework of
QCD-improved factorization. We found that although the twist-3
kaon distribution amplitude dominates the spectator interactions,
it will suppress the decay rates slightly. The weak annihilation
diagrams induced by $(S-P)(S+P)$ penguin operators, which are
formally power-suppressed by order $(\Lqcd/m_b)^2$, are chirally
and logarithmically enhanced. Therefore, these annihilation
contributions are not subject to helicity suppression and can be
sizable. The predicted branching ratio of $B^-\to\phi K^-$ is
$(3.8\pm0.6)\times 10^{-6}$ in the absence of annihilation
contributions and it becomes $(4.3^{+3.0}_{-1.4})\times 10^{-6}$
when annihilation effects are taken into account. The prediction
is consistent with  CLEO and BaBar data but smaller than the BELLE
result.

} \pagebreak

\section{Introduction}
Previously CLEO has put an upper limit on the decay mode $B\to\phi
K$ \cite{Gao}: \be {\cal B}(B^\pm\to\phi K^\pm)<5.9\times 10^{-6}.
\en However, both CLEO \cite{Briere}, BELLE \cite{Chang} and BaBar
\cite{Babar} have recently reported the results:
 \be
 {\cal B}(B^\pm\to\phi K^\pm)=\cases{
(5.5^{+2.1}_{-1.8}\pm0.6)\times 10^{-6} & CLEO, \cr
(7.7^{+1.6}_{-1.4}\pm0.8)\times 10^{-6} & BaBar, \cr
(13.9^{+3.7+1.4}_{-3.3-2.4})\times 10^{-6} & BELLE,} \label{exp1}
 \en
and
 \be
 {\cal B}(B^0\to \phi K^0)=\cases{ (5.4^{+3.7}_{-2.7}\pm 0.7)\times
10^{-6}<12.3\times 10^{-6} & CLEO, \cr
(8.1^{+3.1}_{-2.5}\pm0.8)\times 10^{-6} & BaBar, \cr <16.0\times
10^{-6} & BELLE.} \label{exp2}
 \en

It is known that the neutral mode $B^0\to\phi K^0$ is a pure
penguin process, while the charged mode $\phi K^-$ receives an
additional (though very small) contribution from the tree diagram.
The predicted branching ratio is very sensitive to the
nonfactorizable effects which are sometimes parameterized in terms
of the effective number of colors $N_c^{\rm eff}$; it falls into a
broad range $(13\sim 0.4)\times 10^{-6}$ for $N_c^{\rm
eff}=2\sim\infty$ \cite{CY99}. Therefore, a theory calculation of
the nonfactorizable corrections is urgently needed in order to
have a reliable prediction which can be used to compare with
experiment.

A calculation of $B\to \phi K$ within the framework of
QCD-improved factorization has been carried out recently in
\cite{He}. However, the analysis of \cite{He} is limited to the
leading order in $1/m_b$ and hence the potentially important
annihilation contributions which are power-suppressed in the heavy
quark limit are not included.

In the present paper we will analyze the decay $B\to \phi K$
within the framework of QCD-improved factorization. We will study
the important twist-3 effects on spectator interactions and also
focus on the annihilation diagrams which are customarily assumed
to be negligible based on the helicity suppression argument.
However, weak annihilations induced by the $(S-P)(S+P)$ penguin
operators are no longer subject to  helicity suppression and hence
can be sizable. This is indeed what we found in this work.

\section{Generalized Factorization}
The effective Hamiltonian relevant for $B\to \phi K$ has
the form
\be
{\cal H}_{\rm eff}(\Delta B=1) &=& {G_F\over\sqrt{2}}\Bigg\{
V_{ub}V_{us}^* \Big[c_1(\mu)O_1(\mu)+c_2(\mu)O_2(\mu)\Big] \non \\
&& -V_{tb}V_{ts}^*\left(\sum^{10}_{i=3}c_i(\mu)O_i(\mu)+c_g(\mu)
O_g(\mu)\right) \Bigg\}+{\rm h.c.},
\en
where
\be
&& O_1= (\bar ub)_\vma(\bar su)_\vma, \qquad\qquad\qquad\qquad~~
O_2 = (\bar u_\alpha b_\beta)_\vma(\bar s_\beta u_\alpha)_\vma,
\non \\  && O_{3(5)}=(\bar sb)_\vma\sum_{q'}(\bar
q'q')_{\vma(\vpa)}, \qquad \qquad O_{4(6)}=(\bar s_\alpha
b_\beta)_\vma\sum_{q'}(\bar q'_\beta q'_\alpha)_{ \vma(\vpa)},
\\ && O_{7(9)}={3\over 2}(\bar sb)_\vma\sum_{q'}e_{q'}(\bar
q'q')_{\vpa(\vma)}, \qquad O_{8(10)}={3\over 2}(\bar s_\alpha
b_\beta)_\vma\sum_{q'}e_{q'}(\bar q'_\beta
q'_\alpha)_{\vpa(\vma)}, \non  \\ && O_g=\,{g_s\over
8\pi^2}\,m_b\bar s\sigma^{\mu\nu}G_{\mu\nu}^a {\lambda^a\over 2}\,
(1+\gamma_5)b, \non
\en
with $(\bar q_1q_2)_{_{V\pm A}}\equiv\bar q_1\gamma_\mu(1\pm
\gamma_5)q_2$, $O_3$--$O_6$ being the QCD penguin operators,
$O_{7}$--$O_{10}$ the electroweak penguin operators, and $O_g$ the
chromomagnetic dipole operator.

In the generalized factorization approach for hadronic weak
decays, the decay amplitudes of $B\to\phi K$ read (in units of
$G_F/\sqrt{2}$) \cite{CCTY,AKL}
\begin{eqnarray} \label{famp}
A(B^- \to  K^- \phi)= &&-V_{tb} V^{*}_{ts}\Bigg\{ \left[ a_3 + a_4
+ a_5 -{1\over 2}( a_7+ a_9 + a_{10}) \right] X^{(B^- K^-,\phi)}
\nonumber\\ && + \left[ a_4 + a_{10}- 2(a_6 + a_8) {m^{2}_{B^- }
\over (m_s + m_u)(m_b + m_u)} \right] X^{(B^- ,\phi K^-)} \Bigg\}
\non \\ && + V_{ub} V^{*}_{us} a_1 X^{(B^-,\phi K^-)}, \\ A(\ov
B^0 \to  \ov K^0 \phi)= &&  - V_{tb} V^{*}_{ts} \Bigg\{ \left[ a_3
+ a_4 + a_5 - {1\over 2} (a_7 + a_9 + a_{10} )\right] X^{(\ov B^0
\ov K^0,\phi)}\nonumber\\ && +\left[ a_4 - {1\over 2} a_{10}-
(2a_6 - a_8 ) {m^{2}_{\ov B^0 } \over (m_s + m_d)(m_b + m_d)}
\right] X^{(\ov B^0 ,\phi\ov K^0 )} \Bigg\},  \non
\end{eqnarray}
where the factorized terms
\be
X^{(BK,\phi)} &\equiv & \la \phi| (\bar{s}s)_\vma|0\ra\la
K|(\bar{q}b)_\vma|\ov B \ra=2f_\phi\,m_\phi F_1^{ B
K}(m_{\phi}^2)(\vp^*\cdot p_{_{B}}), \non \\ X^{(B,K\phi)}
&\equiv& \la \phi K|(\bar s q)_\vma |0\ra\la 0|(\bar q b)_\vma
|\ov B\ra=2f_Bm_\phi A_0^{\phi K}(m_B^2)(\vp^*\cdot p_{_{B}}),
\en
can be expressed in terms of the form factors $F_1^{BK}$,
$A_0^{\phi K}$ (for the definition of form factors, see
\cite{BSW}) and the decay constants $f_\phi$ and $f_B$, and the
nonfactorized contributions parametrized in terms of $\chi_i$ are
lumped into the effective number of colors $\nc$:
\be
\left({1\over \nc}\right)_i\equiv {1\over N_c}+\chi_i,
\en
so that the effective parameters $a_i$ appearing in Eq.
(\ref{famp}) read
\be
a_{2i}=c_{2i}+{1\over (\nc)_{2i}}c_{2i-1},\quad\qquad
a_{2i-1}=c_{2i-1}+{1\over (\nc)_{2i-1}}c_{2i}. \label{effai}
\en
It is known that the parameters $a_3$ and $a_5$ depend strongly on
$\nc$, while $a_4$ is $\nc$-stable (see, for example,
\cite{CCTY,AKL}). Therefore, the prediction of $B\to \phi K$ rates
is sensitive to $\nc$ and hence to the nonfactorizable terms
$\chi_i$; it varies from $13\times 10^{-6}$ to $0.4\times 10^{-6}$
for $\nc$ ranging from 2 to $\infty$ \cite{CY99}.

Owing to the unknown form factor $A_0^{\phi K}(m^2_B)$ at large
$q^2$, it is conventional to neglect the annihilation contribution
based on the argument of helicity suppression, which amounts to
having a vanishing form factor $A_0^{\phi K}(m_B^2)$. However,
this argument is valid only for $(V-A)(V-A)$ interactions but not
for $(S-P)(S+P)$ ones. This explains the large enhancement factor
of $m_B^2/(m_b m_s)$ for the penguin contributions [see Eq.
(\ref{famp})]. Therefore, it is conceivable that the annihilation
contribution could be sizable and significant.

\section{Nonfactorizbale effects in Penguin amplitudes}
We next proceed to compute the nonfactorizable effects in the
QCD-improved factorization approach. For simplicity we will
neglect the light quark masses. In the chiral limit, kaon is
massless, but the $\phi$ meson has a finite mass.  We consider the
vertex corrections and hard spectator interactions depicted in
Fig. 1 as well as the annihilation diagrams shown in Fig. 2.
Recently we have analyzed $B\to J/\psi K$ decays within the
framework of QCD factorization \cite{CY00}. The study of $B\to\phi
K$ is quite similar to the $J/\psi K$ mode except for the absence
of weak annihilations in the latter. The reader is referred to
\cite{CY00} for details. The resultant amplitudes are
\begin{eqnarray}  \label{ampqcdf}
A(B^- \to  K^- \phi)= &&-V_{tb} V^{*}_{ts}\Bigg\{ \left[ a_3 + a_4
+ a_5 -{1\over 2}( a_7+ a_9 + a_{10}) \right] X^{(B^- K^-,\phi)}
\nonumber\\ && + \left[ (c_3 +c_9){\cal A}^1_{nf}+(c_5+c_7){\cal
A}_{nf}^2+(c_6 + c_8+{1\over 3}(c_5+c_7)){\cal A}_f
\right]\Bigg\}+ V_{ub} V^{*}_{us} c_2{\cal A}^1_{nf}, \non \\
A(\ov B^0 \to  \ov K^0 \phi)= && - V_{tb} V^{*}_{ts} \Bigg\{
\left[ a_3 + a_4 + a_5 - {1\over 2}
(a_7 + a_9 + a_{10} )\right] X^{(\ov B^0 \ov K^0,\phi)} \\
&& + \left[ (c_3 -{1\over 2}c_9){\cal A}^1_{nf}+(c_5-{1\over
2}c_7){\cal A}_{nf}^2+(c_6 +{1\over 3}c_5-{1\over 2} c_8-{1\over
6}c_7 ){\cal A}_f \right]\Bigg\}, \non
\end{eqnarray}
where
 \be \label{ai}
 a_3 &=& c_3+{c_4\over N_c}+{\alpha_s\over
4\pi}\,{C_F\over N_c} c_4\left[-\left(\matrix{ 18\cr
14\cr}\right)-12\ln{\mu\over m_b}+f_I+f_{II}\right], \non \\
 a_4 &=& c_4+{c_3\over N_c}+{\alpha_s\over 4\pi}\,{C_F\over N_c}
\Biggl\{ c_3\left[-\left(\matrix{ 18\cr
14\cr}\right)-12\ln{\mu\over m_b}+f_I+f_{II}\right]+(c_3-{c_9\over
2}) (G(s_s)+G(s_b)) \non \\ &&-c_1\left({\lambda_u\over
\lambda_t}G(s_u)+{\lambda_c\over\lambda_t}G(s_c)\right)+
\sum_{q=u,d,s,c,b}(c_4+c_6+{3\over2}e_q(c_8+c_{10})) G(s_q)+c_g
G_g\Biggr\}, \non  \\
 a_5 &=&c_5+{c_6\over N_c}-{\alpha_s\over
4\pi}\,{C_F\over N_c} c_6\left[-\left(\matrix{ 6\cr
18\cr}\right)-12\ln{\mu\over m_b}+f_I+f_{II}\right],   \\
 a_7 &=& c_7+{c_8\over N_c}-{\alpha_s\over 4\pi}\,{C_F\over N_c}
c_8\left[-\left(\matrix{
6\cr 18\cr}\right)-12\ln{\mu\over m_b}+f_I+f_{II}\right]-{\alpha\over 9\pi}N_c C_e, \non \\
 a_9 &=& c_9+{c_{10}\over N_c}+{\alpha_s\over 4\pi}\,{C_F\over N_c}
c_{10}\left[-\left(\matrix{ 18\cr 14\cr}\right)-12\ln{\mu\over
m_b}+f_I+f_{II}\right]-{\alpha\over 9\pi}N_c C_e, \non \\
 a_{10} &=& c_{10}+{c_{9}\over
N_c}+{\alpha_s\over 4\pi}\,{C_F\over N_c}
c_{9}\left[-\left(\matrix{ 18\cr 14\cr}\right)-12\ln{\mu\over
m_b}+f_I+f_{II}\right]-{\alpha\over 9\pi} C_e. \non
 \en
In Eq. (\ref{ai}), the upper entry of the matrix is evaluated in
the naive dimensional regularization (NDR) scheme for $\gamma_5$
and the lower entry in the 't Hooft-Veltman (HV) renormalization
scheme, $C_F=(N_c^2-1)/(2N_c)$, $s_q=m_q^2/m_b^2$, $\lambda_{q'}=
V_{q'b}V^*_{q's}$, and $\alpha$ the electromagnetic fine-structure
coupling constant. The other terms in (\ref{ai}) are
 \be
 G(s) &=& {2\over 3}-{4\over 3}\ln{\mu\over m_b}+4\int^1_0 d\xi\,
 \Phi^\phi(\xi)\int^1_0
du\,u(1-u)\ln[s-u(1-u)(1-\xi)], \non
\\ G_g &=& -\int^1_0d\xi\,\Phi^\phi(\xi){2\over 1-\xi}, \non\\
C_e&=& \left({\lambda_u\over
\lambda_t}G(s_u)+{\lambda_c\over\lambda_t}G(s_c)\right)(c_2+{c_1\over
N_c}), \en
 where $\Phi^\phi(\xi)$ is the light-cone distribution
amplitude (LCDA) of the $\phi$ meson, which will be discussed
shortly. $B\to \phi K$ do not receive factorizable contributions
from $a_6$ and $a_8$ except for the annihilation topologies. The
nonfactorizable annihilation contributions ${\cal A}_{nf}^{1,2}$
and the factorizable annihilation amplitude ${\cal A}_f$ will also
be elucidated on below.

\begin{figure}[tb]
\vspace{-4cm}\psfig{figure=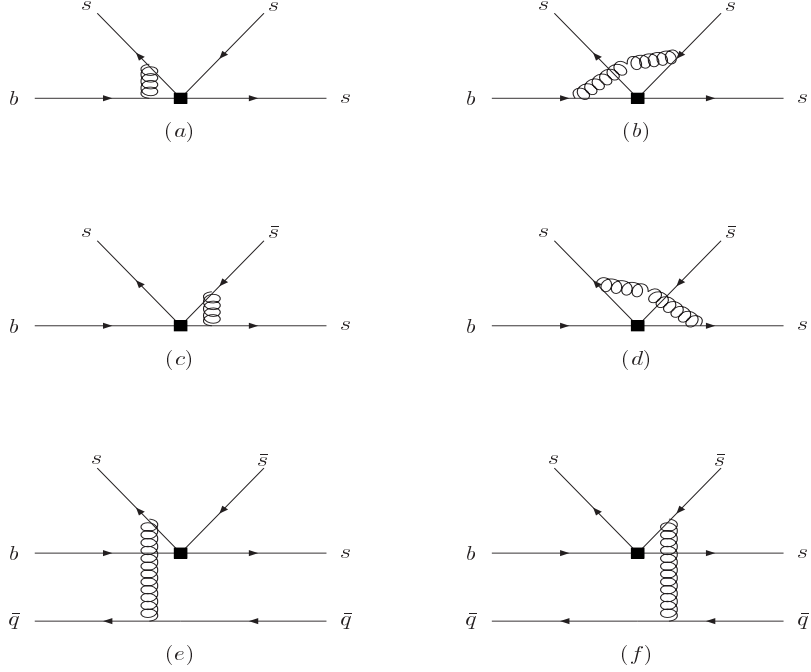,height=7.5in,width=6.5in}
\vspace{-6cm}
    \caption{{\small Vertex and spectator corrections to $B\to \phi K$.}}
   \label{fig:phiK1}
\end{figure}

Note that the effective parameters $a_i$ appearing in Eqs.
(\ref{ai}) are renormalization scale and $\gamma_5$-scheme
independent.  Since only one gluon exchange has been considered in
the annihilation diagrams (see Fig. 2), one may wonder the scale
and scheme dependence of the annihilation amplitude given in Eq.
(\ref{ampqcdf}). Fortunately, as we shall see below, the
annihilation contribution is predominated by the penguin effects
characterized by the parameters $a_6=c_6+{1\over 3}c_5$ and
$a_8=c_8+{1\over 3}c_7$ multiplied by $\mu_\chi$. It turns out
that the scale dependence of $a_6$ and $a_8$ is canceled by the
corresponding dependence in $\mu_\chi$ owing to the runing quark
masses. Consequently, the annihilation amplitude is essentially
scale independent.

The hard scattering kernel $f_I$ appearing in Eq. (\ref{ai}) reads
\be
f_I &=& \int^1_0d\xi\,\Phi^\phi(\xi)\Bigg\{{3(1-2\xi)\over
1-\xi}\ln\xi-3i\pi+{2z(1-\xi)\over 1-z\xi}+3\ln(1-z)  \non \\
&&+\left({1-\xi\over (1-z\xi)^2}-{\xi\over
[1-z(1-x)]^2}\right)z^2\xi\ln z\xi+{z^2\xi^2[\ln(1-z)-i\pi]\over
[1-z(1-\xi)]^2}\Bigg\}, \label{fI} \en where $z\equiv
m^2_\phi/m_B^2$. For completeness, we have included the $\phi$
mass corrections to $f_I$,\footnote{Eq. (\ref{fI}) can be obtained
from Eq. (19) of \cite{Chay} or from Eqs. (2.22) and (2.23) of
\cite{CY00} by neglecting the $\xi^2$ terms arising from the
transverse wave function $\Phi^\phi_\bot$ and applying the
relation
$F_0^{BK}(m_\phi^2)/F_1^{BK}(m_\phi^2)=(m_B^2-m_\phi^2)/m_B^2$ for
form factors.} though such corrections are very small. In the
$m_\phi\to 0$ limit, $f_I$ has the same expression as that in
$B\to \pi\pi$ decay \cite{BBNS1}, as it should be. The hard
scattering kernel $f_{II}$ arises from the hard spectator diagrams
Figs. 1e and 1f and has the form \cite{CY00}
 \be
f_{II} &=& {\alpha_s(\mu_h)\over\alpha_s(\mu)} {4\pi^2\over
N_c}\,{f_K f_B\over F_1^{BK}(m^2_\phi)m_B^2}\,{1\over 1-z}
\int^1_0 {d\drho\over \drho}\,\Phi^B_1(\drho)\int^1_0 {d\xi\over
\xi}\,\Phi^\phi(\xi) \non
\\ && \times\int^1_0 {d\deta\over \deta}
\left(\Phi^K(\deta)+{2\mu_\chi(\mu_h)\over m_B}\, {1\over
(1-z)^2}\,{\Phi^K_\sigma(\deta)\over 6\deta}\right), \label{fII}
 \en
where
 \be
2\mu_\chi(\mu)={2m_K^2\over m_s(\mu)+m_u(\mu)}={-4\la \bar
qq\ra\over f_K^2}  \label{chi}
 \en
is proportional to the quark condensate, the $B$ meson wave
function $\Phi^B_1$ is defined by \cite{BBNS1}
 \be \la 0|\bar q_\alpha(x)b_\beta(0)|\bar
B(p)\ra\!\!\mid_{x_+=x_\bot=0}=-{if_B\over 4}[(p\!\!\!/
+m_B)\gamma_5]_{\beta\gamma}\int^1_0d\drho\, e^{-i\drho
p_+x_-}[\Phi^B_1(\drho)+n\!\!\!/_-\Phi^B_2(\drho)]_{\gamma\alpha},
\label{Bwf}
 \en
with $n_-=(1,0,0,-1)$, and $\Phi^K_\sigma$ is a twist-3 kaon LCDA
defined in the tensor matrix element \cite{Ballp}:
 \be
\la K^-(P)|\bar s(0)\sigma_{\mu\nu}\gamma_5 u(x)|0\ra &=& -{i\over
6}{f_K m_K^2\over
m_s+m_u}\left[1-\left({m_s+m_u\over m_K}\right)^2\right]  \non \\
&& \times(P_\mu x_\nu-P_\nu x_\mu) \int^1_0 d\deta \,e^{i\deta
P\cdot x}\Phi_\sigma^K(\bar \eta).
 \en
Since asymptotically $\Phi^K_\sigma(\deta)= 6\deta(1-\deta)$, the
logarithmic divergence of the $\deta$ integral in Eq. (\ref{fII})
implies that the spectator interaction is dominated by soft gluon
exchanges between the spectator quark and the strange or
anti-strange quark of $\phi$. Hence, QCD factorization breaks down
at twist-3 order. Note that the hard gluon exchange in the
spectator diagrams is not as hard as in the vertex diagrams. Since
the virtual gluon's momentum squared there is $k^2=(-\bar \rho
p_B+\bar\eta p_K)^2\approx -\bar\rho\bar\eta m_B^2\sim \mu_h m_b$,
where $\mu_h$ is the hadronic scale $\sim 500$~MeV, we will set
$\alpha_s\approx \alpha_s(\sqrt{\mu_h m_b})$ in the spectator
diagrams. For the second term in (\ref{fII}), due to the end point
divergence, the scale may correspond to a softer scale $\mu_s$.
However since $\alpha_s\mu_\chi$ is weakly scale-dependent, we can
treat it at the $\sqrt{\mu_h m_b}$ scale. The corresponding Wilson
coefficients in the spectator diagrams are also evaluated at the
$\mu_h$ scale.

The infrared divergence is manifested in the integral $\int_0^1
d\bar\eta/\bar\eta$. However, it is known that the collinear
expansion cannot be correct in the end point region owing to the
transverse momentum $\langle k_T\rangle$ of the quark which is
averagely about 300~MeV, the order of the meson's size. Thus the
lower limit of $\int_0^1 d\bar\eta/\bar\eta$ should be
approximately proportional to $2\langle k_T\rangle/m_b$; or
equivalently, $\int_0^1 d\bar\eta/\bar\eta$ can be approximately
replaced by $\int_0^1 d\bar\eta/(\bar\eta+\langle
2k_T\rangle/m_b$). A consistent treatment of $k_T$ in the
calculation is still an issue since  $k_T$ itself is a higher
twist effect in the QCD factorization approach.  Thus we will
treat the divergent integral as an unknown ``model" parameter and
write
 \be Y\equiv\int^1_0 {d\deta\over \deta}=\ln\left({m_B\over
\mu_h}\right)(1+\rho_H), \label{logdiv}
 \en
with $\rho_H$ being a complex number whose phase may be caused by
soft rescattering \cite{BBNS2}. We see that although the
scattering kernel induced by the twist-3 LCDA of the kaon is
formally power suppressed in the heavy quark limit, it is chirally
enhanced by a factor of $(2\mu_\chi/\Lqcd)\sim {\cal O}(10)$,
logarithmically enhanced by the infrared logarithms.

Finally, we wish to remark that the leading-twist LCDAs of the
$\phi$ meson are given by \cite{Ballv}
\be
\la\phi(P,\lambda)|\bar s(x)\gamma_\mu s(0)|0\ra  &=& f_\phi
m_\phi\,{\vp^{*(\lambda)}\cdot x\over P\cdot x} P_\mu\int^1_0
d\xi\,e^{i\xi P\cdot x}\Phi^\phi_\|(\xi), \non \\
\la\phi(P,\lambda)|\bar s(x)\sigma_{\mu\nu}s(0)|0\ra &=&
-if_\phi^T (\vp^{*(\lambda)}_\mu P_\nu-\vp^{*(\lambda)}_\nu
P_\mu)\int^1_0 d\xi\,e^{i\xi P\cdot x}\Phi^\phi_\bot(\xi),
\label{Jwf}
\en
where $\vp^*$ is the polarization vector of $\phi$, $\xi$ is the
light-cone momentum fraction of the strange quark in $\phi$,
$f_\phi$ and $f^T_\phi$ are vector and tensor decay constants,
respectively, but the latter is scale dependent. Although
$\Phi^\phi_\|$ and $\Phi^\phi_\bot$ have the same asymptotic form,
it is found that the transverse DA does not contribute to $f_I$
and $f_{II}$ if light quarks are massless. The contribution of
$\Phi^\phi_\bot$ to vertex corrections is suppressed by a factor
of $m_\phi/m_B$ and hence can be neglected.

\section{Annihilation amplitudes}
As shown in \cite{BBNS1}, the annihilation amplitude is formally
power suppressed by order $\Lqcd/m_b$. Nevertheless, it has been
stressed in the PQCD approach that annihilation contributions in
hadronic charmless $B$ decays are not negligible \cite{Li}. There
are four weak annihilation diagrams as depicted in Fig. 2. We
first consider the annihilation amplitudes induced by $(V-A)(V-A)$
operators. The first two diagrams, Figs. 2a and 2b, are
factorizable diagrams and their contributions are of order
$m_\phi^2/m_B^2$ and hence can be neglected. Indeed, the
factorizable annihilation amplitude should vanish in $m_\phi\to 0$
limit owing to current conservation. It is easily seen that only
$O_{\rm odd}$ operators contribute to the nonfactorizable
annihilation diagrams Figs. 2c and 2d. It turns out that the
nonfactorizable annihilations are dominated by Fig. 2d owing to an
endpoint contribution. Explicit calculations yield [see Eq.
(\ref{ampqcdf})]:
\be
{\cal A}^1_{nf} &=& -2H\Bigg\{ \int^1_0
d\drho\,d\bar\xi\,d\eta\,{\Phi^B_1(\drho)
\Phi^\phi(\bar\xi)\Phi^K(\eta)\over (\drho-\bar\xi)\bar\xi\eta
}+\int^1_0 d\rho\,d\bar\xi\,
d\eta\,{\Phi^B_1(\rho)\Phi^\phi(\bar\xi)\Phi^K(\eta)\over
\eta[(\rho-\bar\xi)(\rho-\eta)-1] }\Bigg\},  \non
\\ &\cong& 2H\Bigg\{ 6(Y'-1)\int^1_0
d\eta{\Phi^K(\eta)\over \eta}-\int^1_0 d\bar\xi\,
d\eta\,{\Phi^\phi(\bar\xi)\Phi^K(\eta)\over
\eta(\bar\xi\eta-\bar\xi-\eta) }\Bigg\}, \label{A1nf} \en where
\be H={\alpha_s\over 4\pi}\,{C_F\over N_c}\,{4\pi^2\over
N_c}\,{f_B f_K f_\phi m_\phi\over m_B^2}\,(\vp^*\cdot p_{_{B}}),
\en and we have applied the approximation $\rho\approx 1$ and
$\drho=1-\rho\approx 0$. In Eq. (\ref{A1nf}) the first term in
brackets comes from Fig. 2d, while the second term from Fig. 2c.
Since the soft phase of annihilation diagrams is not necessarily
the same as that of the spectator diagram, we write \be
Y'\equiv\int^1_0 {d\deta\over \deta}=\ln\left({m_B\over
\mu_h}\right)(1+\rho_A), \label{logdiv'}
 \en
where the phase is characterized by the complex parameter
$\rho_A$.

\begin{figure}[tb]
\vspace{-4cm}\psfig{figure=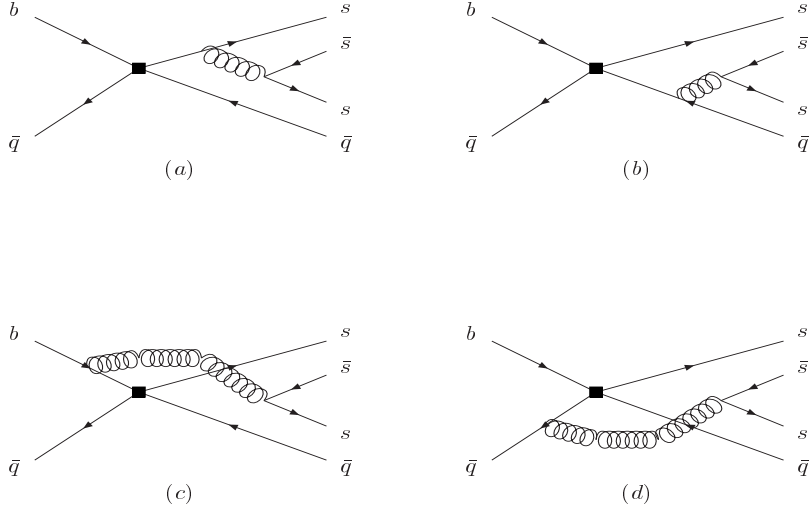,height=7.5in,width=6.5in}
\vspace{-7.6cm}
    \caption{{\small Annihilation diagrams for $B\to \phi K$ decays.}}
   \label{fig:phiK2}
\end{figure}

For $(V-A)(V+A)$ operators $O_5-O_8$, the twist-2 kaon DA makes no
contribution. Therefore, we need to consider the twist-3 kaon
LCDAs $\Phi^K_p$ and $\Phi^K_\sigma$ with the former being defined
in the pseudoscalar matrix element \cite{Ballp}: \be \la
K^-(P)|\bar s(0)i\gamma_5 u(x)|0\ra = {f_K m_K^2\over
m_s+m_u}\int^1_0 d\deta\,e^{i\deta P\cdot x}\Phi_p^K(\deta). \en
The factorizable annihilation amplitude has the expression: \be
{\cal A}_f &=& 4N_c H\left({2\mu_\chi\over m_B}\right)\int^1_0
d\bar\xi\int^1_0d\eta
\,\Phi^\phi(\bar\xi)\Bigg[\Phi^K_p(\eta)\left({1\over
\bar\xi^2}-{1\over 2\bar\xi}\right)\,{1\over
\eta}+\Phi^K_\sigma(\eta)\,{1\over 4\bar\xi \eta^2}\Bigg], \non\\
&=& 24N_c H\left({2\mu_\chi\over m_B}\right)Y'(Y'-{1\over 2}),
\en
where we have applied the LCDAs:
$\Phi^\phi(\bar\xi)=6\bar\xi(1-\bar\xi)$, $\Phi^K_p(\eta)=1$ and
$\Phi^K_\sigma(\eta)=6\eta(1-\eta)$. Likewise, the nonfactorizable
annihilations induced by the penguin operators $O_5$ and $O_7$
read
\be
{\cal A}^2_{nf} = -H\left({2\mu_\chi\over m_B}\right)\Bigg\{
6Y'(Y'-1) +\int^1_0 d\bar\xi\int^1_{\Lambda_{_{\rm QCD}}\over m_b}
d\eta\,\Phi^\phi(\bar\xi)\Phi^K_p(\eta)\,{2-\eta\over
\bar\xi\eta\,(\bar\xi\eta-\bar\xi-\eta) }\Bigg\},
\en
where the dominated first term in brackets stems from Fig. 2d.
Note that we have introduced a cutoff $\Lqcd/m_b$ to regulate the
infrared divergence occurred in the second term.

Although the annihilation amplitudes ${\cal A}_f$ and ${\cal
A}_{nf}^2$ are formally of order $(\Lqcd/m_b)^2$, they receive two
large enhancements: one from the chiral enhancement
$(2\mu_\chi/\Lqcd)\sim {\cal O}(10)$ and the other from the
logarithmic endpoint divergence of the infrared divergent integral
$Y'$. Consequently, the annihilation effects can be sizable.
Physically, this is because the penguin-induced annihilation
contributions are not subject to helicity suppression.

\section{Results and Discussions}
To proceed for numerical calculations, we employ the meson LCDAs
as follows:
\be
\Phi^K(\deta,\mu^2) &=& 6\deta(1-\deta)\left(1+\sum_{n=1}^\infty
a_{2n}^K(\mu^2)C_{2n}^{3/2}(2\deta-1)\right), \non \\
\Phi^B_1(\drho) &=& N_B\drho^2(1-\drho)^2{\rm exp}\left[-{1\over
2}\left({\drho m_B\over \omega_B}\right)^2\right], \\
\Phi^\phi(\bar\xi) &=& \Phi^\phi_\|(\bar\xi) =6\bar\xi(1-\bar\xi),
\non \en where $C_n^{3/2}$ are Gegenbauer polynomials and the
values of the Gegenbauer moments $a_n^K$ are available in
\cite{Ballp}, $\omega_B=0.25$ GeV, and $N_B$ is a normalization
constant. We use the decay constants $f_K=0.16$ GeV, $f_B=0.19$
GeV, $f_\phi=0.237$ GeV, and the running quark masses:
$m_b(m_b)=4.40$ GeV, $m_s(m_b)=90$ MeV $m_d(m_b)=4.6$ MeV and
$m_u(m_b)=2.3$ MeV. The next-to-leading-order Wilson coefficients
$c_i(\mu)$ in NDR and HV $\gamma_5$-schemes are taken from Table
XXII of \cite{Buras96}; they are evaluated at $\mu=\ov
m_b(m_b)=4.40$ GeV and $\Lambda^{(5)}_{\ov{\rm MS}}=225$ MeV. For
form factors we use $F_1^{BK}(m_\phi^2)=0.38$ as a benchmarked
value. Note that $F_1^{BK}(m_\phi^2)=0.407$ in the
Bauer-Stech-Wirbel model \cite{BSW}, while it is 0.37 in a QCD sum
rule calculation \cite{qcdsr}.

For the parameters $\rho_H$ in (\ref{logdiv}) and $\rho_A$ in
(\ref{logdiv'}), in principle they may be complex due to
final-state soft rescattering.  We find that the decay rate is
much more sensitive to $\rho_A$ than to $\rho_H$. Presumably, some
information on the parameter $\rho$ can be extracted from the
study of $B\to K\pi$ modes. It has been shown recently in
\cite{BBNS2} that increasing the parameter $|\rho_A|$ from 1 to 2
would increase the corresponding error on the $K\pi$ branching
ratios in which case it would require considerable fine-tuning of
the strong interaction phase of $Y'$ in annihilation diagrams to
reproduce the experimental value of the branching ratio. Hence, it
is reasonable to assume that the model parameters are in the range
$|\rho|\leq 1$. Writing $\rho_A=|\rho_A|\exp(i\delta)$, the
branching ratio of $B\to\phi K$ vs. the phase $\delta$ is plotted
in Fig. 3. We obtain
 \be
 {\cal B}(B^-\to\phi K^-)=\,(4.3^{+3.0}_{-1.4})\times
 10^{-6},\qquad {\cal B}(B^0\to\phi K^0)=\,(4.0^{+2.9}_{-1.4})\times
 10^{-6},
 \en
where the central value corresponds to the default values
$\rho_A=\rho_H=0$ and the errors come from the variation of
$|\rho_H|$ and $|\rho_A|$ from 0 to 1; that is, the theoretical
uncertainties come from power corrections of twsit-3 spectator
interactions and annihilation contributions. Therefore, the
predicted branching ratio is consistent with CLEO and BaBar
numbers, but smaller than the BELLE result [see (\ref{exp1}) and
(\ref{exp2})]. The corresponding absolute ratio of the
annihilation to penguin amplitudes depends on the annihilation
phase and is at most of order $0.25$. In the absence of
annihilation effects, the branching ratios are given by
 \be
{\cal B}(B^\pm\to\phi K^\pm) &=& \left({F_1^{BK}(m_\phi^2)\over
0.38}\right) (3.8\pm 0.6)\times 10^{-6}, \non \\ {\cal
B}(B^0\to\phi K^0) &=& \left({F_1^{BK}(m_\phi^2)\over 0.38}\right)
(3.6\pm0.6)\times 10^{-6},
 \en
where the error arises from the variation of $\rho_H$ from 0 to 1.

Needless to say, the major theoretical uncertainty stems from the
unknown model parameters $\rho_H$ and $\rho_A$. It should be
stressed that the infrared divergence here is always of the
logarithmic type and other possible linear divergence occurred in
annihilation diagrams with twist-3 wave functions are explicitly
canceled out. As stressed in passing, the infrared divergence
stems from the misuse of the collinear expansion in the end point
region where the effect of the quark's transverse momentum is
important. Since $k_T$ is a higher-twist effect in QCD
factorization, at present we treat the infrared divergent integral
in a model manner.

\begin{figure}[tb]
\psfig{figure=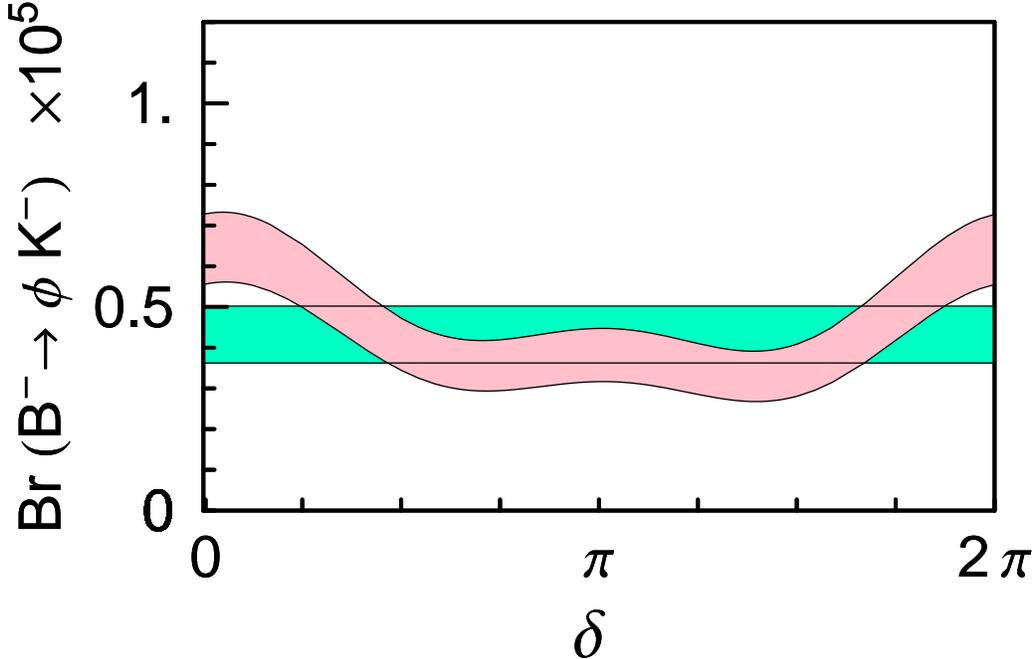}\vskip0.5cm
    \caption{{\small Branching ratio of $B^-\to\phi K^-$ vs. the phase of the
    complex parameter $\rho_A$ [see Eq. (\ref{logdiv'})],
    where the dark (horizontal) and light bands correspond to
    $|\rho_A|=0,~1$, respectively, with the variation of $|\rho_H|$ from 0 to 1.}}
   \label{fig:phiK3}
\end{figure}

Several remarks are in order. (i) The calculations are rather
insensitive to the unitarity angle $\gamma$ as the CKM matrix
element $V_{ub}V^*_{us}$ is considerably suppressed. (ii) The
scattering kernel $f_{II}$ is dominated by the twist-3 effect.
However, since the Wilson coefficients $c_{2i}$ and $c_{2i-1}$
have opposite signs, it turns out that the magnitudes of
$a_{3-10}$ [see Eq. (\ref{ai})] are slightly reduced by the
twist-3 terms and therefore the branching ratio is suppressed by
about 10\% in the presence of twist-3 effects in spectator
interactions. (iii) In the QCD factorization approach, the strong
phase of the annihilation amplitude is of order $\alpha_s^2$ since
it comes from the annihilation diagrams in which the gluon line is
inset with an enclosed quark loop, resembling the
vacuum-polarization bubble. Consequently, the phase of the
annihilation contribution is likely dominated by the soft one
induced by soft scattering as characterized by the parameter
$\rho_A$. This is in contrast to the PQCD approach where the
annihilation contributions have large strong phases \cite{Li}.

\section{Conclusions}
We have analyzed the decay $B\to \phi K$ within the framework of
QCD-improved factorization and taken into account some of
power-suppressed corrections. Our conclusions are:
\begin{enumerate}
\item  Although the twist-3 kaon distribution amplitude dominates
the spectator interactions, it will suppress the decay rates of
$B\to\phi K$ slightly by about 10\%. In the absence of
annihilation contributions, the branching ratio is $(3.8\pm
0.6)\times 10^{-6}$ for $\phi K^-$ and $(3.6\pm 0.6)\times
10^{-6}$ for $\phi K^0$.
\item The weak annihilation diagrams induced by $(S-P)(S+P)$ penguin
operators, which are formally power-suppressed by order
$(\Lqcd/m_b)^2$, are chirally and logarithmically enhanced.
Therefore, these annihilation contributions are not subject to
helicity suppression and in principle can be sizable.
\item The branching ratio is predicted to be
$(4.3^{+3.0}_{-1.4})\times 10^{-6}$ for $B^-\to\phi K^-$ and
$(4.0^{+2.9}_{-1.4})\times 10^{-6}$ for $B^0\to\phi K^0$, where
theoretical uncertainties come from power corrections of twist-3
spectator interactions and annihilation contributions. The
corresponding absolute ratio of annihilation to penguin amplitudes
depends on the annihilation phase and is at most $25\%$.

\end{enumerate}

\vskip 2 cm  \acknowledgments  One of us (H.Y.C.) wishes to thank
Physics Department, Brookhaven National Laboratory for its
hospitality. This work was supported in part by the National
Science Council of R.O.C. under Grant Nos. NSC89-2112-M-001-082
and NSC89-2112-M-033-014.

%%%%%%%%%%%%%%%%%%%%%%%%%%%%%%%%%%%%%%%%%%%%%%%%%%%%%%%%

\end{document}